\documentclass[11pt]{article}
\usepackage{moriond,epsfig}

\def\Journal#1#2#3#4{{#1} {\bf #2}, #3 (#4)}

\def\PLB{{\em Phys. Lett.}  B}
\def\PRL{\em Phys. Rev. Lett.}

\def\PRC{{\em Phys. Rev.} C}

\def\be{\begin{equation}}
\def\ee{\end{equation}}
\def\bea{\begin{eqnarray}}
\def\eea{\end{eqnarray}}

\begin{document}
\vspace*{4cm}
\title{$\pi^{+} \Xi^{-}$ CORRELATIONS AND
THE $\Xi^{*}(1530)$ PUZZLE.}\footnote{Contribution to the proceedings of the XLIVth Rencontres de Moriond
     QCD, La Thuile, Italy, March 2009.}

\author{B.O. Kerbikov}
\address{Institute for Theoretical and Experimental Physics,
117218 Moscow, Russia }

\author{R.Lednicky}
\address{Joint Institute for Nuclear Research, Dubna, Moscow Region, 141980, Russia}

\author{L.V. Malinina}
\address{M.V. Lomonosov Moscow State University, D.V. Skobeltsyn
Institute of Nuclear Physics, 119992, Moscow, Russia }

\author{P. Chaloupka and M. \v{S}umbera,}
\address{Nuclear Physics Institute, Academy of Sciences of the
Czech Republic, 250 68 \v{R}e\v{z}, Czech Republic}

\maketitle\abstracts{
The analysis of the preliminary RHIC data on $\pi^{+}-\Xi^{-}$ correlation function
is carried out. The $\Xi^*(1530)$ resonance is reasonably described.
The value of the fireball radius has been estimated and the sensitivity to
the $\pi^{+}-\Xi^{-}$
$S$-wave  scattering lengths has been tested.
}



Measurement of momentum correlations of the two
low relative momentum particles
produced in heavy ion collisions is an important method to study
the spatio-temporal picture of the emission source at the level of fm $=10^{-15}$ m.
This type of analysis acquired the name of femtoscopy and has been reviewed in e.g.
\cite{pod89}$^-$\cite{lis05}.

At the early stages the studies were focused on the production
of identical pions, since then, measurements have been performed for different systems
of both identical and non-identical hadrons,
high-statistics data sets were accumulated in heavy ion experiments at
AGS, SPS, RHIC accelerators \cite{E895}$^-$\cite{STAR}.
Correlations are significantly affected by the
Coulomb and/or strong final state interactions (FSI) between outgoing particles.
The non-identical particles correlations due to FSI provide information not
only about space-time characteristics of the emitting source, but also about
the average relative space-time separation asymmetry between the emission
points of the two particle species in the pair rest frame \cite{LLED}.
Maybe the most exotic system studied recently by STAR collaboration
is $\pi-\Xi$ \cite{Sumbera07,Chaloupka_SQM06}:
the particles composing the pair have one order of magnitude
difference in mass plus  $\Delta$B=1/$\Delta$S=2 gap in baryon/strangeness
quantum numbers. It is challenging to study
FSI of such exotic meson-baryon system and to extract information
about the $\pi-\Xi$ S-wave scattering lengths.
The other important reason to study  $\pi-\Xi$
correlations is that
multistrange baryons are expected to
decouple earlier, than other particle species
because of their small hadronic cross-sections~\cite{Bass},
allowing one to extract the space-time interval between the
different stages of the fireball evolution.

Preliminary results for the $\pi \Xi$
system are available from STAR Collaboration
\cite{Sumbera07,Chaloupka_SQM06}. The following important observations
were made:
\begin{itemize}
\vspace*{ -.3cm} \item Decomposition of the correlation function
$C(\mathbf{k})\equiv C(k,\cos\theta,\varphi)$ from $10\%$
of the most central Au+Au collisions into spherical harmonic, provided
the first preliminary values of $R=(6.7 \pm 1.0)$ fm and
$\Delta_{out}=(-5.6 \pm 1.0)$ fm. The negative value of the shift
parameter $\Delta_{out}$ indicates that the average emission point of
$\Xi$ is positioned more to the outside of the fireball than
the average emission point of the pion.

\vspace*{ -.3cm}\item  In addition to the Coulomb interaction seen
in previous non-identical particle analysis the $\pi^{+} \Xi^{-}$
correlations at small relative momenta provide sufficiently clear
signal of the strong FSI that reveals itself in a peak
corresponding to the $\Xi^*(1530)$ resonance. The
peak's centrality dependence shows a high sensitivity to the
source size.

\vspace*{ -.3cm}\item Comparison with the FSI model \cite{Pratt-Petriconi} confirms that
theoretical calculations in the Coulomb region are in a
qualitative agreement with the data. They however over-predict
the peak in the $\Xi^*(1530)$-region.

\end{itemize}
Below we present the first results of our calculations of the
$\pi^{+} \Xi^{-}$ FSI and make the comparison with the experimental data.
The problem of FSI in the $\pi^{+} \Xi^{-}$ system is highly intricate since one has to take into
account the following factors \cite{Pratt-Petriconi}$^-$\cite{Led08}:
\begin{itemize}
\vspace*{ -.3cm}\item The superposition of strong and Coulomb interactions
\vspace*{ -.3cm}\item The presence of $\Xi^*(1530)$ resonance
\vspace*{ -.3cm}\item The spin structure of the w.f. including spin-flip.
\vspace*{ -.3cm}\item The fact that the $\pi^{+} \Xi^{-}$ state is a superposition of $I=1/2$
and $I=3/2$ isospin states and that $\pi^{+} \Xi^{-}$
state is coupled to the $\pi^{0} \Xi^{0}$  and that the thresholds of the two channels are non-degenerate.
\vspace*{ -.3cm}\item The contribution from inner potential region where the structure of the strong
interaction is unknown.
\end{itemize}

The outgoing multichannel wave functions (w.f.'s)
 of $\pi^{+} \Xi^{-}$ system $\Psi^{(-)}(\vec{k},\vec{r})$ enter as building blocks into the
correlation function (CF \cite{Pratt-Petriconi}$^-$\cite{Led08}:
\begin{equation}
\label{CF}
 C(\vec{k}) = \sum_{i}\int d\vec{r}S_{i}(\vec{r})|\Psi^{(-)}_{i}(\vec{k},\vec{r})|^{2}\doteq
\sum_{i}\int d\vec{r}S(\vec{r})|\Psi^{(-)}_{i}(\vec{k},\vec{r})|^{2}  ,
\end{equation}
here $\vec{k}$ is a relative momentum of the pair, $S(\vec{r})$
is a universal source function. The out-state w.f.'s $\Psi^{(-)}_{i}(\vec{k},\vec{r})$
have the asymptotic form

\begin{equation}
\label{Psim}
\Psi_{i}^{(-)}(\vec{k},\vec{r}) \simeq  e^{i \vec{k_1} \vec{r} } \delta_{i1}+
f^{*}_{i1}(-\cos{\theta}) \frac{e^{-i k_i r}}{r} \left ( \frac{\mu_i}{\mu_1} \right)^{1/2},
\end{equation}
where $\mu_{i}$ is the reduced mass
(or, for relativistic particles, reduced energy)
of the particles in channel $i=1-4$, ($i=\pi^{+}\Xi^{-},\pi^{0} \Xi^{0}$
without and with the spin flip).

We considered the
Gaussian (in the pair rest frame) model for the source
function:
\begin{equation}
\label{SF}
S(\vec{r})=(8\pi^{3/2}R^{3})^{-1}\exp(-r^{2}/4R^{2}).
\end{equation}
The low energy region of  $\pi \Xi$ interaction up to the $\Xi^*(1530)$ resonance is dominated by $S$- and $P$-waves.
Therefore the w.f. contains two phase shifts with $I=1/2,3/2$ for $S$-wave and four phase shifts with
$I=1/2,3/2$ and $J=1/2,3/2$ for $P$-wave ($J=l \pm 1/2$ is the total momentum).
To reduce the number of parameters we have assumed that the dominant interaction in
$P$-wave occurs in a state with $J=3/2$, $I=1/2$ containing the
$\Xi^*(1530)$ resonance. Since the parameters of $\Xi^*(1530)$ are known from the experiment we are
left with two $S$-wave phase shifts which are expressed in terms of the two scattering lengths
$a_{\rm 1/2}$ and $a_{\rm 3/2}$ with isospin $I=1/2$ and $I=3/2$ correspondingly.

Leaving the technical details for the future full-size publication we present the resulting
expression for the sum of the squares of w.f.'s in Eq.~(\ref{CF}).

\begin{eqnarray}
\sum_{i=1}^{4}|\Psi^{(-)}_{i}(\vec{k}, \vec{r})|^{2}  = | \Psi_{Coul}^{*}(-\vec{k}, \vec{r})+
T_{0}\varphi_{0} Y_{0}^{0}+\frac{2}{\sqrt{3}}T_{1}\varphi_{1} Y_{1}^{0}|^{2} + |\sqrt{\frac{2}{3}}T_{1}\varphi_{1} Y_{1}^{1*}|^{2}+
 \nonumber \\
+\frac{k_2}{k_1} \left( |R_0 \chi_0 Y_{0}^{0} + \frac{2}{\sqrt{3}}R_{1}\chi_{1} Y_{1}^{0}|^{2}
+ | \sqrt{\frac{2}{3}} R_{1}\chi_{1} Y_{1}^{1*}|^{2} \right),
\label{Psi4}
\end{eqnarray}
here $\Psi_{Coul}$ is the pure Coulomb w.f., $k_1$ and $k_2$ are the c.m. momenta in $\pi^{+} \Xi^{-}$
and $\pi^{0} \Xi^{0}$  channels;
the spherical harmonics $Y_{l}^{m}=Y_{l}^{m}(\pi-\theta,\phi+\pi)$
correspond to the reversed direction of the vector $\vec{k}$, the functions
\begin{eqnarray}
\varphi_{l}(\eta_1,\rho_1)=\sqrt{4 \pi} (-i)^{l} e^{-i \sigma_{l}(\eta_1)}H_{l}^{(-)}(\eta_1,\rho_1)/\rho_1
 \nonumber \\
\chi_{l}(\eta_1,\rho_2)=
\left (\mu_i/\mu_1\right)^{1/2}\sqrt{4 \pi} (-i)^{l} e^{-i \sigma_{l}(\eta_1)}H_{l}^{(-)}(0,\rho_2)/\rho_2,
\end{eqnarray}
where $H_{l}^{(-)}(\eta,\rho)$ is a combination of the regular and singular Coulomb functions $F_l$
$G_l$ at a given orbital angular momentum $l$
with the asymptotics $H_{l}^{(-)}(\eta,\rho) \to e^{-i(\rho+\sigma_l-\eta ln2 \rho-l \pi/2)}$,
$\rho_{1}=k_{1}r$, $\rho_{2}=k_{2}r$, $\eta_1=(a_{1}k_{1})^{-1}$, $a_{1}=-214$~fm is a Bohr radius of the
$\pi^{+} \Xi^{-}$ system taking into account the negative sign of the Coulomb repulsion.
The quantities $T_{l}=k_{1} f_{l}^{J;11*}$ and $R_{l}=-(k_{1}k_{2})^{1/2} f_{l}^{J;21*}$ contain the elastic
$(1\to 1)$ and inelastic $(1\to 2)$ scattering amplitudes $f_l^{J;11}$ and $f_l^{J;21}$ at a given total and
orbital angular momentum $J$ and $l$. For the $S$-waves ($l=0$, $J=1/2$), they are expressed through
the scattering lengths $a_{\rm 1/2}$ and $a_{\rm 3/2}$ is a similar way as in pion-nucleon scattering
(see, e.g. \cite{LL98,Led08}).
For the resonance $\Xi^*(1530)$ $P$-wave,
\begin{eqnarray}
T_{1}= -\frac{\Gamma_{1}/2}{E-E_{0}-i\Gamma/2},~~
R_{1} = - \frac{(\Gamma_{1}\Gamma_{2})^{1/2}/2}{E-E_{0}-i\Gamma/2},
\label{Yl}
\end{eqnarray}
where $\Gamma = \Gamma_{1}+\Gamma_{2}$,
$\Gamma_{1} \doteq 2 \Gamma/3$, $\Gamma_{2} \doteq \Gamma/3$.

Expression (\ref{Psi4})
describes the region $r > \epsilon \sim 1$~fm where the strong potential is assumed to vanish.
In the inner region $r<\epsilon$, we substitute Eq.~(\ref{Psi4}) by
$|\Psi_{Coul}|^{2}$ and take into account the effect of strong interaction in a form of a correction
\cite{Pratt-Petriconi}$^-$\cite{Led08}
which depends on the strong interaction time (expressed through the phase shift derivatives) and can be calculated
without any new parameters unless the $S$-wave effective radii are extremely large. It is important that the complete
CF does not depend on $\epsilon$ provided the source function is nearly constant in the region
$r< \epsilon$ \cite{LL82,LL98}.

Fig.~1 presents the results of calculations and the experimental data from
\cite{Sumbera07,Chaloupka_SQM06}.
The solid curve corresponds to the source size
$R = 7.0$~fm and zero $S$-wave scattering lengths.
The results are however practically the same even for the $S$-wave scattering
lengths of the order of one fm. We may conclude that at present experimental errors,
the CF at $R > 7$~fm is practically independent of the $S$-wave scattering parameters.

Similar to the FSI model \cite{Pratt-Petriconi}, our calculations are in agreement with the data in the low-$k$
Coulomb region. Contrary to this model, they are however much closer to the experimental peak in the
$\Xi^*(1530)$ region though, they still somewhat overestimate this peak.
The predicted peak is however expected to decrease due to a strong angular asymmetry of a more realistic
source function obtained from Blast-wave like simulations.

In summary, using a simple Gaussian model for the source function, we have
reasonably described the experimental data on the $\pi^{+} \Xi^{-}$  CF,
estimated the emission source radius and tested the sensitivity to the
low energy parameters of the strong interaction.

\begin{figure}
\begin{center}
\begin{tabular}{c}
\includegraphics[width=10cm]{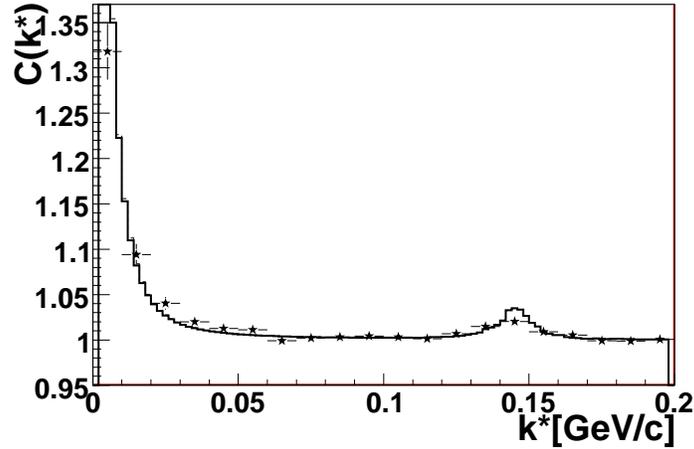}
\end{tabular}
\end{center}
\caption{
The CF of  $\pi^{+} \Xi^{-}$ system for $R=7.0$~fm and zero scattering lengths
(solid line), the experimental data points are from the STAR collaboration.
\label{fig:CF}
}
\end{figure}

\section{Acknowledgments}
This work was partially supported by Russian Foundation for Basic Research
(grants No 08-02-91001 and No 08-02-92496) and
by the GDRE: European
Research Group comprising IN2P3/CNRS, Ecole des Mines de Nantes,
Universite de Nantes, Warsaw University of Technology, JINR Dubna,
ITEP Moscow and Bogolyubov Institute for Theoretical Physics NAS of
Ukraine
and
by the Grant Agency of
the Czech Republic, Grant 202/07/0079 and by Grant LC 07048 of the
Ministry of Education of the Czech Republic.

\end{document}